\documentclass{phb-proc4-auth}
\usepackage{graphicx}
\usepackage{amssymb}
\begin{document}
\begin{frontmatter}
\journal{SCES '04}
\title{NMR study of electronic state in CePt$_3$Si}
\author{Koh-ichi Ueda},
\author{Kouji Hamamoto},
\author{Takao Kohara},
\author{Gaku Motoyama} and
\author{Yasukage Oda}
\address{Graduate School of Material Sci., Univ.of Hyogo, Kamigori-cho, Ako-gun, Hyogo 678-1297, Japan}
%
%
%
%
\begin{abstract}
In this article, we report the temperature dependence of spin-lattice relaxation rates at two Pt sites and one Si site in CePt$_3$Si with a non-centrosymmetric structure center.
$1/T_1$ for both Pt sites between 2 K and 300 K and $1/T_1$ of Si above 3 K might be explained by the contributions from the low-lying crystal-electric-field level and the quasiparticle due to the hybridization between the ground state and conduction electrons.
Just below $T_{\rm c}$ no remarkable enhancement in $1/T_1$ was observed.
The estimated value of superconducting gap is about $2\Delta = 3k_{\rm B}T_{\rm c}$.
\end{abstract}
%
%
\end{frontmatter}
%
%
%
%
%
Recently, Bauer {\it et al.} reported a new heavy fermion superconductor CePt$_3$Si with N\'{e}el temperature ($T_{\rm N}$) and superconducting (SC) transition temperature ($T_{\rm c}$) of 2.2 K and 0.75 K, respectively.\cite{bau}
The first feature of this compound is no spatial inversion center in the chemical unit cell, which favors spin singlet pairing in the superconductivity. 
The second feature is the large ${\rm d}H_{\rm c2}/{\rm d}T$ (= -- 8.5 T/K) and $H_{\rm c2}$ ($\sim $5 T), which suggest that Cooper pairs form out of the heavy quasi-particle state. 
The large $H_{\rm c2}$, which exceeds the estimated Pauli-Clogston limiting field, might be a signature for spin triplet pairing. 
A mixed spin singlet and triplet pairing state might be one of the candidate to answer this paradox.\cite{fri}
To investigate the electronic state of this compound microscopically, we performed $^{29}$Si and $^{195}$Pt NMR experiments on non-annealed and annealed CePt$_3$Si samples, hereafter referred to ``as cast" and ``annealed", respectively.

The starting materials of samples are Ce, Pt and Si with 99.9\%(3N), 99.95\%(3N5) and 6N purity, respectively.
Polycrystalline samples were prepared from stoichiometric amounts of starting materials by arc melting in Ar atmosphere.
The half of ``as cast" sample was annealed successively at 950$^{\rm \circ}$C for 1 week. 
The powder X-ray diffraction measurement showed no extra phase in both ``as cast" and ``annealed" samples. 

Shown in fig. 1 is the NMR spectrum of $^{195}$Pt for aligned powder of CePt$_3$Si, which is applied by the external field parallel to the $c$-axis. 
Under high magnetic field, powder of CePt$_3$Si can be oriented to the magnetically easy-axis ($c$-axis) by some mechanical vibrations due to the difference of parallel and perpendicular susceptibilities to the applied field.
There exist two crystallographically inequivalent Pt sites in CePt$_3$Si: 
one is Pt(1) site which is located near face-centered position of $ac$ plane in the unit cell, the other site Pt(2) is on $c$-axis together with Si site. 
The site occupancy-ratio for Pt(1) and Pt(2) is 2:1. 
From the intensity ratio of them observed at 4.2 K, the larger peak at higher field and the smaller peak at lower field are found to come from Pt(1) and Pt(2) sites, respectively.
A weak shoulder on the spectrum observed at higher field might be due to an incompleteness of alignment by the external field.
Below $T_{\rm N}$, however, their line widths for both Pt(1) and (2) sites become broader due to the internal field from antiferromagnetically ordered Ce magnetic moment. 
The bottom spectrum displays the overlapping NMR lineshape from both Pt sites at 1.4 K. 
Above 60 K, two peaks of Pt(1) and Pt(2) sites become closer each other, and then make a single peak above 80 K (not shown in the figures). 
On the other hand, $^{29}$Si NMR line shape stays sharp and symmetric above 4.2 K. 
No significant difference in the NMR spectra was observed between ``as cast" and ``annealed" samples in the whole $T$ range.

Fig. 2 shows the temperature ($T$) dependence of spin-lattice relaxation rate ($1/T_1$) for Si and Pt sites measured at peak positions in the NMR spectra. 
According to the recent neutron experiment,\cite{met} due to the CEF the six-fold degenerate 4f-levels are split into three sets of doublets, a ground state with first and second excited levels, 1 and 24 meV, respectively.
The relaxation process may be described by the low lying crystal-electric-field level and the quasi particle due to the hybridization between ground state and conduction electrons.
In this case two contributions from the first and second excited levels make the relaxation behavior complicated.
Above 80 K, two relaxation rates for both sites coincide each other, associated with the overlapping NMR line shapes. 
An increase of $1/T_1$ above 100 K may be ascribed to the second excitation level ($\Delta$ =280 K) of CEF measured by the recent neutron scattering experiment.\cite{met}

As shown in this figure, $1/T_1$ for Si in ``as cast" is markedly enhanced with a distinct peak near the antiferromagnetic (AF) transition, reflecting a critical slowing down of the fluctuations of Ce moment. 
However, no peak in $1/T_1$ for ``annealed" sample was observed, as displayed in the inset. 
This sample dependence is consistent with the result of susceptibility measurement: 
only ``as cast" sample has a rapid increase of susceptibility just above $T_{\rm N}$.[5]
Below $T_{\rm N}$, $1/T_1$ for both samples decrease rapidly probably due to the formation of AF gap at the Fermi surface.
Since the overlapping for two peaks of Pt(1) and Pt(2) is seen in fig. 1, it is difficult to measure the relaxation rates of Pt(1) and Pt(2) separately below $T_{\rm N}$. 
So, the $T$ dependence of $1/T_1$ was mainly measured at a peak position in the spectrum.
Below $T_{\rm N}$, $1/T_1$ decreases rapidly and then it becomes proportional to $T$, which means that the system is in the Fermi liquid state. 

According to the published data[1], the SC transition $T$ is estimated as $T_{\rm c}(H)$=0.6 K under a magnetic field of 1.4 T, where our NMR measurements were performed. 
Moreover, the coherence length in the present $T$ range and magnetic field is also estimated as 160~\AA.\cite{clo}
Considering the lattice parameters, the volume-ratio of vortex core in the total volume is estimated as $\sim$50\%.
 Thus, most of relaxation behaviors have the contributions from the normal fluxoid cores via the thermal fluctuation of fluxoids and the spin diffusion to vortex cores. 
In such a case, $1/T_1$ decreases in proportion to $T$ through $T_{\rm c}$. 
The long component is, of course, ascribed to the intrinsic relaxation in SC region. 
So, only the long component, which has a minor part in the nuclear recovery curve, was tentatively evaluated.
 As a result, no remarkable Hebel-Slichter coherence peak was observed just below $T_{\rm c}$ within an experimental error in the measurement, and $1/T_1$ decreases with decreasing $T$. 
Assuming the SC gap is isotropic, the value of SC gap is roughly estimated as about $2\Delta = 3k_{\rm B}T_{\rm c}$. 
%
%
%
%

Note added. - After submission of this paper, no appreciable decrease of $^{29}$Si Knight shift was observed for parallel direction to the $c$-axis below $T_{\rm c}$. This indicates that spin triplet pairing is probably realized for superconductivity. Detailed data will be published elsewhere in the very near future.
%
%
\begin{figure}
     \centering
     \includegraphics[width=45mm]{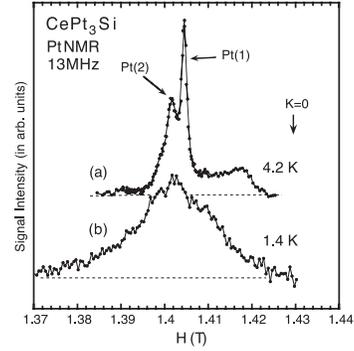}
     \caption{Pt NMR spectrum of (a) magnetically oriented powder sample to magnetic field direction at 4.2 K, and (b) overlapping Pt spectra obtained in oriented powder sample at 1.4 K.} 
 \end{figure} 
 \begin{figure}
     \centering
     \includegraphics[width=65mm]{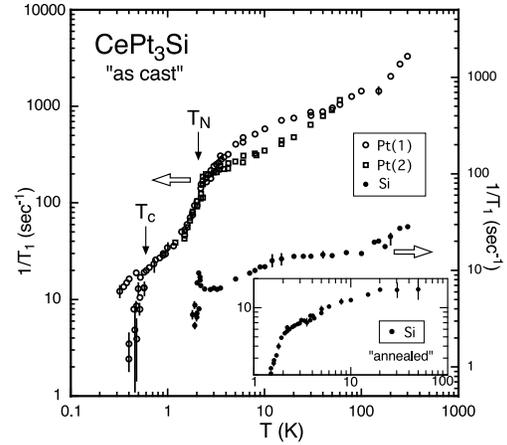}
     \caption{Temperature dependence of spin-lattice relaxation rate ($1/T_1$) of Pt(1) site, Pt(2) site and Si site in ``as cast" sample. Inset is that of Si site in ``annealed" sample.} 
 \end{figure}  
%
%
\end{document}